\documentclass[prl,twocolumn,showpacs]{revtex4-1}
\usepackage{amssymb,amsfonts,amsmath} 
\usepackage{graphicx}
\usepackage{bm}
\usepackage{color}
\usepackage{hyperref}

\begin{document} 

\title{Spectral properties of one-dimensional Fermi systems after an interaction quench}

\author{D.M.\ Kennes}  
\affiliation{Institut f{\"u}r Theorie der Statistischen Physik, RWTH Aachen University 
and JARA---Fundamentals of Future Information
Technology, 52056 Aachen, Germany}

\author{C.\ Kl\"ockner}  
\affiliation{Institut f{\"u}r Theorie der Statistischen Physik, RWTH Aachen University 
and JARA---Fundamentals of Future Information
Technology, 52056 Aachen, Germany}

\author{V.\ Meden} 
\affiliation{Institut f{\"u}r Theorie der Statistischen Physik, RWTH Aachen University and 
JARA---Fundamentals of Future Information
Technology, 52056 Aachen, Germany}

\begin{abstract} 

We show that the single-particle spectral properties of gapless one-dimensional Fermi systems 
in the Luttinger liquid state reached at intermediate times after 
an abrupt quench of the two-particle interaction are highly indicative of 
the unusual nonequilibrium nature of this state. The line shapes of the momentum 
integrated and resolved spectral functions strongly differ from 
their ground state as well as finite temperature equilibrium counterparts. Using 
an energy resolution improved version of radio-frequency spectroscopy 
of quasi one-dimensional cold Fermi gases it should be possible to experimentally 
identify this nonequilibrium state by its pronounced spectral signatures.  
  
\end{abstract}

\date{\today} 
\maketitle

Closed one-dimensional (1d) Fermi systems are promising candidates to realize unusual 
nonequilibrium steady states. In many 1d out of equilibrium systems 
integrals of motion beyond energy conservation hinder that observables and 
the reduced density matrix of subsystems relax to values expected 
for one of the canonical ensembles. Even for {\em generic} models, in which 
those are absent, the restricted phase space available for scattering at least delays
{\em thermalization}. An intensively studied question is under which conditions 
1d systems prepared in the ground state of an 
initial Hamiltonian $H_{\rm i}$ and brought out of equilibrium by time-evolving them with 
$H_{\rm f}$ of similar form  but with changed parameters 
eventually thermalize \cite{Cazalilla06,Rigol07,Kollath07,Manmana07,Kollar08,Barthel08,Mitra11,Karrasch12,Mitra12,Mitra13,Sirker14,Essler14}. 
Quench protocols of this type can experimentally be 
realized in cold atomic gases \cite{Bloch08,Polkovnikov11}. 
Increasing theoretical evidence was gathered 
that in an extended time regime following the after-quench transient one a quasi 
stationary state exists which does show nonthermal behavior \cite{Manmana07,Karrasch12,Mitra13,Kennes13,Essler14}.   
Here we are interested in the properties 
of this {\em intermediate time steady state} regardless if it is the true steady state or 
if further relaxation sets in on a well separated larger time scale \cite{Mitra13}.
For nongeneric models which are {\em mappable on free ones} 
the system is stuck in this state
for all further times; it can be described by a generalized Gibbs ensemble (GGE) 
\cite{Cazalilla06,Rigol07,Kollar08,Iucci09,Kennes10,Calabrese11,Essler12,Fagotti13,Sotiriadis14}.
If the same holds for more complex models 
which cannot directly be linked to noninteracting ones but are 
characterized by an infinite set of local conservation laws is heavily 
investigated \cite{Rigol07,Fagotti14,Wouters14,Mierzejewski14,Pozsgay14}.

\begin{figure}[t]
\includegraphics[width=0.9\linewidth,clip]{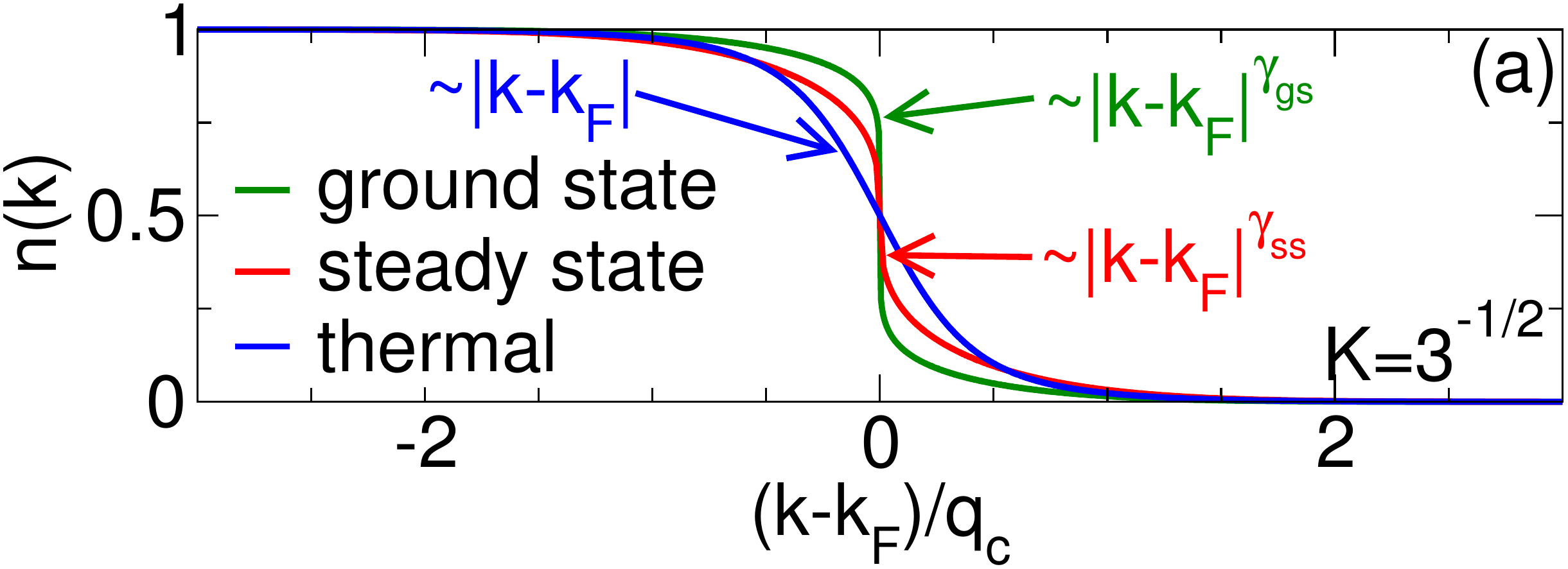}
\includegraphics[width=0.92\linewidth,clip]{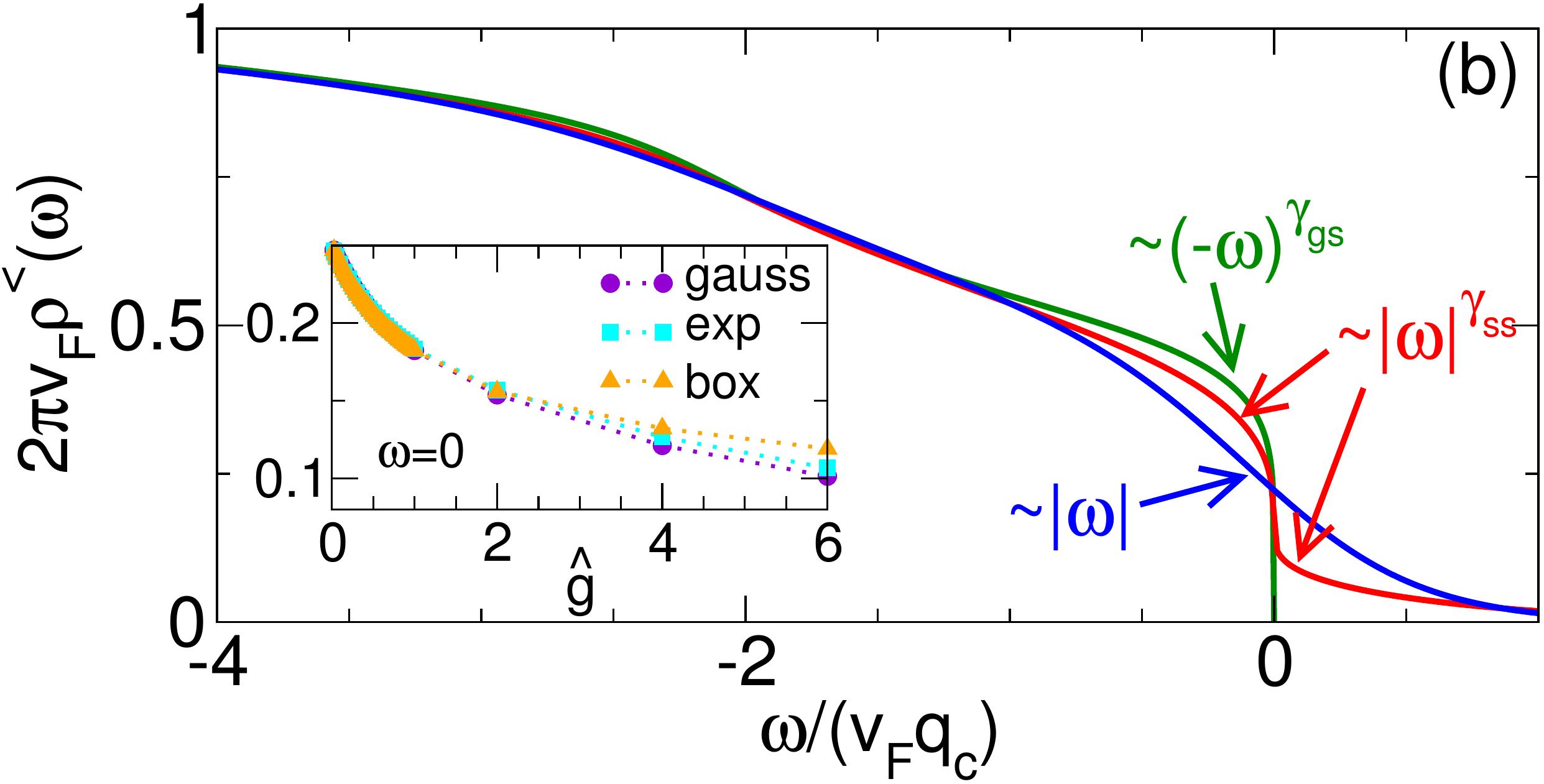}\vspace*{-0.2cm}
\caption{(Color online) (a) The TL model momentum distribution functions of 
the ground state, the intermediate time nonequilibrium steady state following the 
interaction quench with $\hat g = 2$,
and the thermal equilibrium state with temperature $T/(v_{\rm F}q_{\rm c}) \approx 0.2804$ 
for which the average energy is equal to the one quenched into the system.
(b) The corresponding momentum integrated spectral functions. 
The smooth features at $\omega/(v_{\rm F} q_{\rm c})$ of order 1 are 
nonuniversal and depend on the chosen potential.
The inset of (b) shows the nonuniversal $2 \pi v_{\rm F}\rho_{\rm ss}^<(0)$ as a function of the 
interaction $\hat g$ for the Gaussian potential, an exponential one $\hat g(q) = \hat g e^{-|q/q_{\rm c}|}$, 
and a box potential $\hat g(q) = \hat g \Theta(q_{\rm c} - |q|)$.  }
\label{fig1}
\end{figure}

\begin{figure}[t]
\includegraphics[width=1\linewidth]{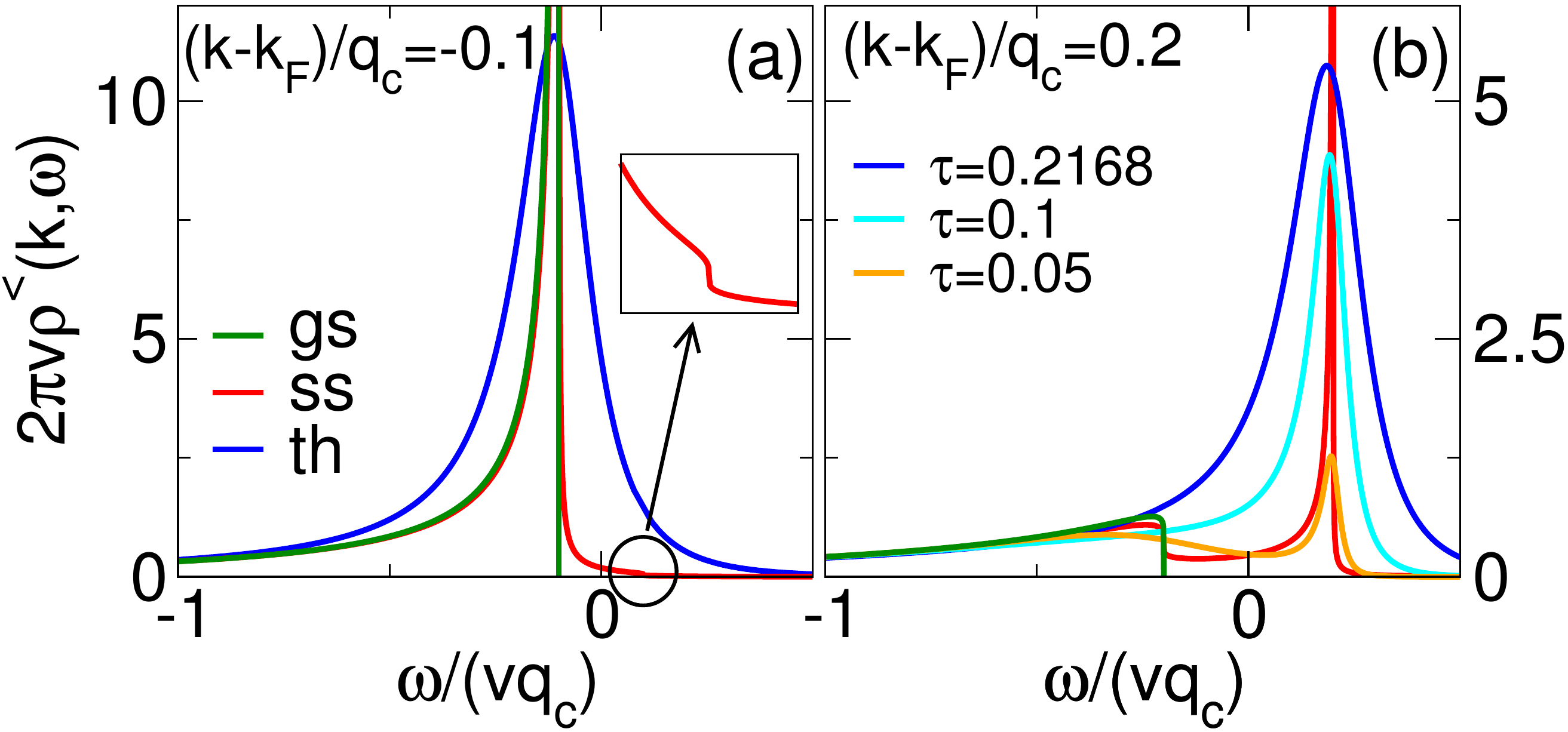}\vspace*{-0.2cm}
\caption{(Color online) 
The momentum resolved spectral functions of the TL model in 
the ground state, the intermediate time nonequilibrium steady state following the 
interaction quench with $\hat g = 2$,
and the thermal equilibrium state with temperature $\tau=T/(vq_{\rm c}) \approx 0.2168$ 
for which the average energy is equal to the one quenched into the system. 
The steady state function shows a characteristic peak at positive energies for $k-k_{\rm F}>0$. 
For this case thermal curves with two additional $\tau$ are shown for comparison.}
\label{fig2}
\end{figure}

Already in {\em equilibrium} 1d metals are peculiar. They fall into the Luttinger 
liquid (LL) universality class which is characterized by power-law decay of 
correlation functions with interaction dependent exponents \cite{Giamarchi03,Schoenhammer05}. 
The Tomonaga-Luttinger model (TLM), a field theoretical model for which the exact 
computation of correlation functions is straightforward using bosonization, 
constitutes the low-energy fixed point model under a renormalization group (RG) 
procedure of all microscopic models falling into the LL class \cite{Giamarchi03,Schoenhammer05}. 
The power-law decay of the TLM's ground state lesser Green function $i G^<(x,t')=
\left< \psi^{\dag}(0,0) \psi^{}(x,t')\right>$, with the field operator $\psi$ in the Heisenberg 
picture, leads to a power-law nonanalyticity of the momentum distribution function 
$n(k) = \int dx e^{-ikx} i 
G^<(x,0)$ for momenta close to the Fermi momentum $k_{\rm F}$ (instead of a jump 
by the quasi-particle weight in Fermi liquids; see Fig.~\ref{fig1}(a)).
For $\omega \to 0^-$, that is close to the chemical potential, the spectral 
function $\rho^<(\omega) = \int dt' e^{i \omega t'} i G^<(0,t')/(2 \pi)$ approaches zero 
following a power law and vanishes for $\omega>0$ (see Fig.~\ref{fig1}(b)).  
The line shape of the momentum resolved spectral function 
$\rho^<(k,\omega) = \int dx \int dt'  e^{-i(kx-\omega t')}  i G^<(x,t')/(2 \pi)$ strongly 
differs from the one of a Fermi liquid (see Fig.~\ref{fig2}). In electronic systems
$\rho^<(k,\omega)$ and $\rho^<(\omega)$ can directly be measured using 
photoemission spectroscopy; for a recent such measurement on a LL see e.g.~\cite{Blumenstein11}.

Recent studies of {\em microscopic lattice models} in their metallic phase 
indicate that the intermediate time nonequilibrium steady state as well 
as the time evolution towards it shows LL properties when performing a 
{\em global quench of the two-particle 
interaction.} The results obtained for $n(k)$ \cite{Karrasch12}, the Friedel oscillations of the 
density, as well as the  static density response \cite{Kennes13} were consistent with 
typical LL power laws derived for the TLM in the steady state reached after the same 
type of interaction quench. 
Compared to their ground state counterparts the dependencies of the exponents 
of static correlation functions of the TLM on the LL parameter $K$ are modified 
\cite{Cazalilla06,Iucci09,Uhrig09,Rentrop12}. 
E.g. for a spinless model the densities' Friedel oscillations off an inhomogeneity decay with exponent 
$-K$ in the interacting ground state but with $-(K^2+1)/2$ in the steady state 
for the initial state being the noninteracting ground state \cite{Kennes13}. For weak 
interactions $K = 1 - U/U_{\rm c} + {\mathcal O}(U^2)$, with the amplitude $U$ 
of the interaction and a model dependent $U_{\rm c}$. Thus both 
exponents agree to leading order in $U$. 
Furthermore, for a given experimental system $K$ is usually unknown. It is therefore
rather challenging to distinguish the ground and the 
nonequilibrium steady state based on measurements of static correlation functions. 
We show that the {\em single-particle  spectral properties} derived from {\em dynamical} 
intermediate time steady state correlation functions \cite{Essler12} in contrast significantly differ 
from their ground state counterparts. In particular, 
$\rho^<(\omega)$ has  {\em finite spectral weight at} $\omega \geq 0$ (see Fig.~\ref{fig1}(b)) 
with a {\em power-law} approach of the nonuniversal constant $\rho^<(0)$ from both sides.
Even more strikingly $\rho^<(k,\omega)$ after the quench 
shows a prominent {\em peak at positive frequencies} when $k> k_{\rm F}$ while 
the ground state has vanishing spectral weight for 
$\omega \geq 0$ (see Fig.~\ref{fig2}(b)). The spectral 
functions of the intermediate time steady state can as well easily be distinguished from 
finite temperature $T$ spectra (see Figs. \ref{fig1}(b) 
and \ref{fig2}). Our results are obtained for the TLM using bosonization and supported by 
microscopic lattice model calculations based on analytical arguments and the numerical 
density-matrix RG (DMRG) \cite{Schollwoeck11}. 

Cold Fermi gases confined to 1d provide the most promising systems to realize the 
intermediate time nonequilibrium LL steady state. In such the single-particle 
spectral functions can be measured by radio-frequency spectroscopy \cite{Stewart08,Gaebler10,Feld11}. 
After improving the energy resolution by roughly one order of magnitude 
it should be possible to detect the unique spectral features of the nonequilibrium LL 
state of correlated matter.   

\textit{The TLM and bosonization}---After linearizing the dispersion relation of the 
spinless 1d Fermi gas around $\pm k_F$ its Hamiltonian for a translational invariant chain 
of length $L$ can be written in bosonized form as \cite{Giamarchi03,Schoenhammer05}
\begin{eqnarray}
H = \sum_{n>0}  \left[ k_n \left( v_{\rm F} + \frac{g(k_n)}{2 \pi}\right) \left( b^\dag_n b^{}_n + 
b^\dag_{-n} b^{}_{-n} \right) \right. \nonumber \\ 
 \left. +  k_n   \frac{g(k_n)}{2 \pi} \left( b^\dag_{-n} b^\dag_{n} + 
b^{}_{-n} b^{}_{n} \right) \right] ,
\label{TLHam}
\end{eqnarray}
with bosonic ladder operators $b^{(\dag)}_n$ associated to the densities of left- and right-moving
fermions, the Fermi velocity $v_{\rm F}$, and $k_n = 2 \pi n/L$, $n \in {\mathbb Z}$. For simplicity---but 
not affecting our main results---we assume that the intra- and interbranch 
two-particle potentials are equal; $g(q)$ falls off on a scale $q_{\rm c}$. Employing a Bogoliubov transformation
Eq.~(\ref{TLHam}) can be written as $H=\sum_{n \neq 0} \omega(k_n) \alpha_n^\dag \alpha_n^{} + 
E_{\rm gs}$ with new bosonic ladder operators $\alpha_n^{(\dag)}$ of dispersion $\omega(k) = v_{\rm F} |k| 
\sqrt{1+\hat g(k)}$, where $\hat g = g/(\pi v_{\rm F})$. The ground state $\left| {\rm gs} \right>$ 
is the product of the 
vacua with respect to the $\alpha_n$ and the ground state energy $E_{\rm gs}= -2 \sum_{n \neq 0} 
\omega(k_n) s^2(k_n)$, with $s^2(k) = \{ [1+\hat g(k)/2]/\sqrt{1+\hat g(k)} -1 \}/2$ \cite{Rentrop12}.

Applying bosonization of the field operator \cite{Giamarchi03,Schoenhammer05} 
computing the lesser Green function $i G_t^<(x,t')=
\left< \psi^{\dag}(0,0) \psi^{}(x,t')\right>_{\rho(t)}$ of the right moving fermions 
with the statistical operator $\rho(t)$ obeying the von 
Neumann equation is straight forward. The initial conditions considered are: $\rho_{\rm i}=
\left| {\rm gs} \right> \left< {\rm gs} \right|$, leading to the $t$ independent ground state Green function, 
$\rho_{\rm i} = \left| {\rm gs}_0 \right> \left< {\rm gs}_0 \right|$, with the 
noninteracting ground state $\left| {\rm gs}_0 \right>$ (the vacua product of the $b_n$), leading to 
the $t$ {\em dependent} Green function of the quenched system, and $\rho_{\rm i} = \exp{(- H/T)}/{\mathcal Z}$ with 
the (canonical) partition function ${\mathcal Z}$, leading to the temperature $T$ equilibrium
Green function ($t$ independent). 
The time evolution is performed with $H$ Eq.~(\ref{TLHam}). Up to a momentum sum  
$G_t^<(x,t')$ can be given in closed form \cite{SM}. In those expressions the thermodynamic 
limit can be taken. In the quench protocol the time $t$ can 
subsequently be sent to infinity. As the TLM can be written in free bosons the resulting
steady state is the state of our interest. We have explicitly verified that the corresponding 
{\em dynamical} correlation function 
$i G_{\rm ss}^<(x,t')$ can as well be obtained directly by computing $\left< \psi^{\dag}(0,0) 
\psi^{}(x,t')\right>_{\rho_{\rm GGE}}$ with the GGE density matrix $\rho_{\rm GGE} = 
\exp{(-\sum_{n \neq 0} \lambda_n \alpha_n^\dag \alpha_n^{})}/Z_{\rm GGE}$, the Lagrange
multipliers $\lambda_n$ fixed such that $\left< {\rm gs}_0 \right| N_n 
\left| {\rm gs}_0 \right> = \left< N_n\right>_{\rho_{\rm GGE}}$ 
for the conserved eigenmode occupancies $N_n = \alpha_n^{\dag} \alpha_n^{}$, and taking 
$L \to \infty$ \cite{Essler12}.
To obtain $n(k)$ and $\rho^<(\omega)$ we numerically perform the momentum integral
as well as the corresponding Fourier integral. We assume $\hat g(q) = \hat g 
e^{-[(q/q_{\rm c})^2/2]}$ but note that our conclusions are independent of the details
of the $q$ dependence as long as $g(q=0) < \infty$, i.e.~the interaction is 
(sufficiently) short ranged (screened) in real space.  
For $|k-k_{\rm F}|/q_{\rm c}, 
|\omega|/(v_{\rm F} q_{\rm c}) \ll 1$, analytical results can be derived. 
To obtain $\rho^<(k,\omega)$ we instead resort to an often applied 
ad hoc procedure in which $\omega(k)$ is replaced by its lowest order expansion 
$v k$ with the renormalized velocity $v=v_{\rm F} \sqrt{1+\hat g(0)}$ and the 
momentum integral is regularized in the ultraviolet by multiplying by 
$e^{-|q/q_{\rm c}|}$. The consequences of this {\em approximation} \cite{Meden99}
are discussed below. Then the $q$ integral can be performed 
analytically \cite{SM} and the remaining Fourier integrals numerically.

In Fig.~\ref{fig1}(a) we compare the ground state, the steady state as well as the thermal $n(k)$. 
In the latter $T$ is chosen such that the average energy corresponds to 
the energy quenched into the system (see also Figs.~\ref{fig1}(b) and \ref{fig2}). 
The ground and steady states are characterized
by power laws $|1/2-n(k)| \sim (|k-k_F|/q_{\rm c})^\gamma$ with $\gamma_{\rm gs} = (K+K^{-1}-2)/2$ 
\cite{Giamarchi03,Schoenhammer05} and $\gamma_{\rm ss} = (K^2+K^{-2}-2)/4$ \cite{Cazalilla06,Uhrig09,Rentrop12},
respectively, where $K^{-1}= \sqrt{1+ \hat g(0)}$. 
This holds for $\gamma < 1$ and independent of the form of $\hat g(q)$; 
for $\gamma >1$ \cite{Meden99} as well as for $T>0$ \cite{Schoenhammer93,Karrasch12a} 
the leading dependence 
for $k \approx k_{\rm F}$ is linear. 
As the analytical expression of the two exponents in terms of $K$
differ, one is tempted to conclude that measuring $n(k)$ allows to distinguish the 
LL ground and the steady states. However, for an experimental system $K$ is usually 
unknown and measuring a single exponent does thus not allow to conclude which of the states
is realized. Furthermore, for small interactions the exponents only differ by a factor of 
two \cite{Rentrop12} and even if $K$ would be known within some bounds from the measurement
of another observable 
distinguishing the two states would require a very precise determination of an exponent which 
is hardly achievable. Similar reasoning holds for other static correlation functions 
(see the introduction) \cite{Iucci09,Karrasch12,Kennes13}. We next show 
that the {\em spectral properties} are more suitable to distinguish the two states. 

In Fig.~\ref{fig1}(b) we compare the three $\rho^<(\omega)$. 
In the ground state $\rho_{\rm gs}^< \sim [-\omega/(v_{\rm F} q_{\rm c})]^{\gamma_{\rm gs}}$ 
\cite{Giamarchi03,Schoenhammer05} (regardless of the size of $\gamma_{\rm gs}$ \cite{Meden99}; 
compare to $n_{\rm gs}(k)$). $G_{\rm gs}^<(0,t')$ is analytical 
in the upper-half of the complex $t'$ plane thus $\rho^<_{\rm gs}(\omega)$ vanishes for 
$\omega > 0$---in equilibrium ($T=0$) only the occupied states are visible in 
photoemission. As discussed in \cite{Schoenhammer93} for $|\omega|/(v_{\rm F} q_{\rm c}), 
T/(v_{\rm F} q_{\rm c})  \ll 1$ one finds $\rho_{\rm th}^< \approx (c_1 \omega^2 + c_2 T^2)^{\gamma_{\rm gs}/2} 
f(\omega,T)$, with the Fermi function $f$ and (dimensionful) constants $c_{1/2}$. For fixed 
$T$ and $|\omega|/T \ll 1$ the corrections to the finite weight at $\omega=0$ are thus linear 
in $\omega$. Both, the vanishing of the weight in the ground state for $\omega \geq 0$ as well 
as the linear behavior close to $\omega=0$ in thermal equilibrium are in stark contrast to
our results for $\rho_{\rm ss}^<(\omega)$. As shown in Fig.~\ref{fig1}(b) $\rho^<_{\rm ss}$ 
has weight for all $\omega$ and is finite at $\omega=0$. The zero frequency weight is 
nonuniversal; it does not only depend on $\hat g(0)$ (as $K$, and thus $\gamma$, as well as $v$ do) 
but on the details of the $q$ dependence of the potential (inset of Fig.~\ref{fig1}(b)).  
$\rho^<_{\rm th} (0)$ increases with increasing $T$. If $T$ is chosen such that the average 
energy corresponds to the energy quenched into the system, $T$, and thus  $\rho^<_{\rm th} (0)$, 
increases with $\hat g$. In contrast, $\rho^<_{\rm ss} (0)$ is a decreasing function of $\hat g$. 
Using asymptotic analysis we find \cite{SM} $|\rho^<_{\rm ss}(\omega) 
- \rho^<_{\rm ss}(0)| \sim |\omega|^{\gamma_{\rm ss}}$ for $\omega \to 0^-$ {\em and} $\omega \to 0^+$
(if $\gamma_{\rm ss} <1$; else $\rho^<_{\rm ss}(\omega) - \rho^<_{\rm ss}(0) \sim -\omega$). 
In short, $ \rho^<_{\rm ss}(0)$ is finite and nonuniversal, but the corrections to it show 
universal (`critical') power-law scaling. This rather unusual behavior is very indicative of 
the intermediate time nonequilibrium LL steady state. 

An even more easily observable 
hallmark of the nonequilibrium state is found in $\rho^<_{\rm ss}(k,\omega)$.
The $k-k_{\rm F} < 0$ spectral functions of Fig.~\ref{fig2}(a) are all dominated by a peak 
located at $\omega = v (k-k_{\rm F})$. While $\rho^<_{\rm gs}$ vanishes 
for $\omega> v (k-k_{\rm F})$, $\rho^<_{\rm ss}$ carries weight also for these energies. 
In fact, a weak second structure is visible at $\omega = -v(k-k_{\rm F})$ (inset of Fig.~\ref{fig2}(a)). 
Due to the outlined ad hoc procedure the features at $\omega = \pm v (k-k_{\rm F})$ are power laws with 
exponents given in \cite{SM}. We expect that without the approximation 
underlying this procedure the spectra look very similar but are most likely not characterized by 
algebraic singularities as was discussed for the ground state in \cite{Meden99}. The dominating 
peak is broadened in the finite $T$ canonical ensemble. The most striking difference
between the ground and intermediate time steady state spectra is found for $k-k_{\rm F}>0$ (Fig.~\ref{fig2}(b)). 
In this case $\rho_{\rm gs}^<$ only shows a weak bump around $\omega = - v(k-k_{\rm F})$ and vanishes for 
larger energies, while the dominating feature of $\rho_{\rm ss}^<$ is a narrow peak at  
{\em positive energies} $\omega =  v(k-k_{\rm F})$; within the 
ad hoc procedure it is given by a power law \cite{SM}. The peak is much more narrow 
than the broad one found in $\rho^<_{\rm th}$. To further emphasize the differences of $\rho^<_{\rm ss}$ and 
$\rho^<_{\rm th}$ in Fig.~\ref{fig2}(b) we show the latter for two additional $T$ smaller than the one 
determined by the quenched energy. To aquire a physical understanding of the positive 
energy peak we first discuss the threshold behavior of $\rho_{\rm gs}^<$. The ground state 
corresponds to the $\alpha$-boson vacuum. For fixed $k-k_{\rm F} >0$ one has to accomodate a momentum
$-(k-k_{\rm F})$ which can be done by creating $\alpha$-bosons with momenta 
$q \in {\mathbb R}$ summing up to this value. The lowest energy state with $|\omega| = v (k-k_{\rm F})$ 
is the one in which a single boson with momentum $-(k-k_{\rm F})$ is created leading to the threshold. 
In contrast, the steady and the thermal states already contain $\alpha$-bosons. It is thus possible 
to {\em destroy} $\alpha$-bosons with positive momentum. Since the single boson state with $\omega =   
v (k-k_{\rm F})$ carries the largest weight \cite{Schoenhammer05} a peak at this energy emerges. The 
difference in the width follows from the different distributions of $\alpha$-bosons.

{\em Lattice models and the inverse quench}---We next investigate if the revealed spectral 
characteristics of the intermediate time nonequilibrium LL state can also be found in 
microscopic models and are thus universal. Analytical \cite{Fagotti14,Wouters14,Pozsgay14} 
or semi-analytical \cite{Kennes13} approaches to the (intermediate time) steady state 
applicable to microscopic models which are not directly linked to free ones are rare. 
Thus one frequently resorts to 
numerics and results can only be obtained up to a certain time or for small systems (restricting 
the time for which the results are of relevance due to recurrence effects). The computation 
of spectral functions would require to perform {\em two} 
time evolutions up to sizable times to reach the intermediate time steady state (intermediate $t$) and 
to be able to perform the Fourier transform with sufficient energy resolution  (large $t'$). For a 
general setup this is currently out of reach. We thus resort to a particular one, which, however, 
shows the spectral characteristics discussed above. 

\begin{figure}[t]
\includegraphics[width=1\linewidth]{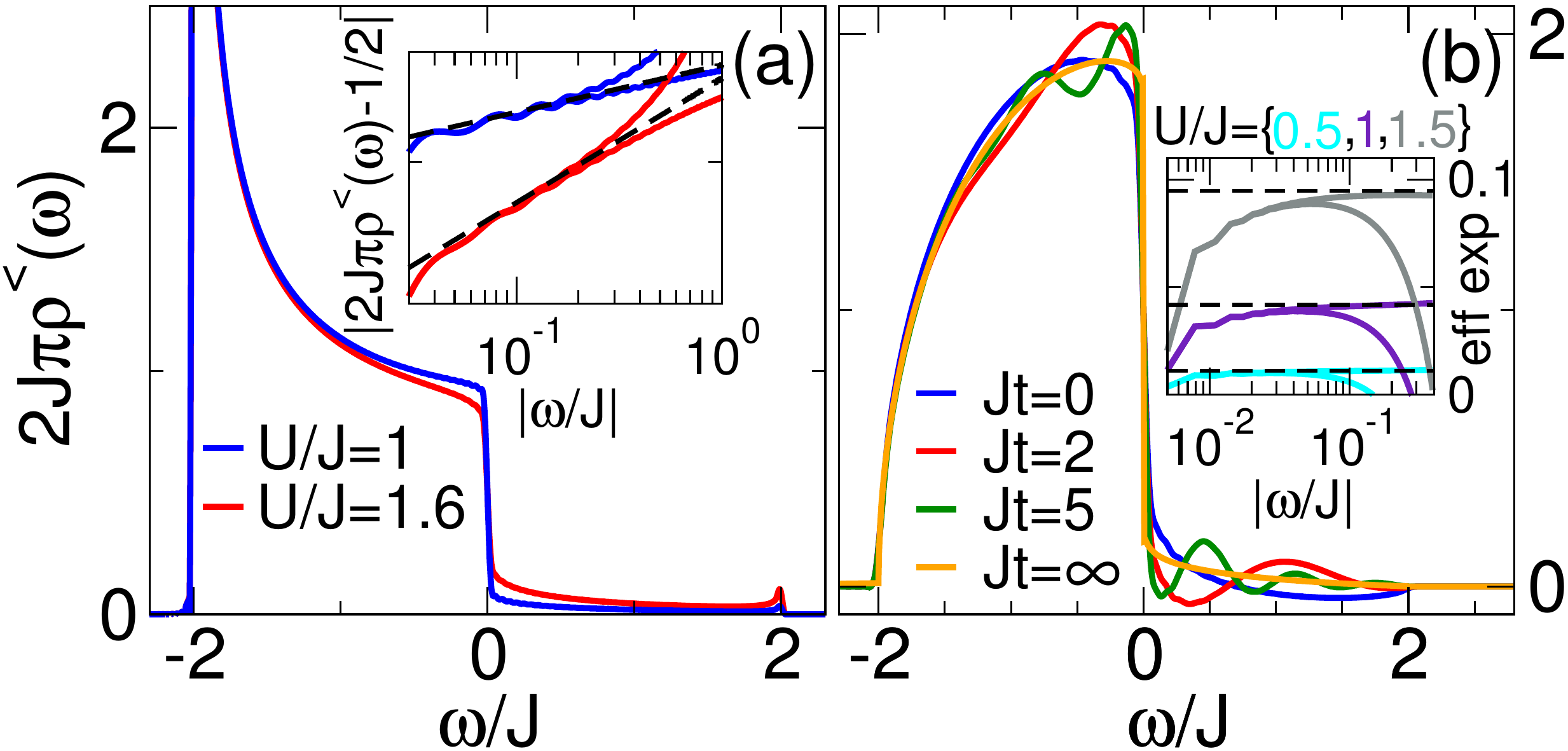}\vspace*{-0.2cm}
\caption{(Color online) 
The momentum integrated spectral function for the lattice model
of spinless fermions with nearest neighbor interaction $U$ after the 
inverse quench. (a) $\rho^<_{\rm ss}(\omega)$ of the translational invariant model. 
Inset: $|\rho^<_{\rm ss}(\omega)-\rho_{\rm ss}^<(0)|$ on a log-log scale 
indicating the power-law behavior with $\gamma_{\rm gs}$ for $|\omega| \to 0$. 
The dashed lines are the expected power laws with $K$ taken from Bethe 
ansatz \cite{Giamarchi03,Schoenhammer05}. The upper bound over which the 
power law manifests depends on $U$ \cite{Karrasch12a} and 
${\rm sgn} (\omega)$. At small $|\omega|$ the power law is cut off by $t'^{-1}$. 
(b) Spectral function at the first site $\rho_t^{<}(1,\omega)$ for 
open boundaries with $U/J=1$ at different $t$. Inset: 
$d \ln|\rho^<_{\rm ss}(1,\omega)-\rho_{\rm ss}^<(1,0)|/d \ln|\omega|$, that is the 
effective exponent, in the steady state. The dashed lines indicate the expected exponents} 
\label{fig3}
\end{figure}

Instead of performing the above quench we consider the {\em inverse} one by starting in the 
{\em interacting} ground state and {\em time evolving with the noninteracting 
Hamiltonian.} Exploiting translational invariance it is straight forward to show 
that $G_t^<(x,t')$ of an arbitrary model does not depend on $t$ 
and is given by $i G_{\rm ss}^<(x,t') = \sum_{k} e^{-i(\varepsilon_k t'+ kx)} 
n_{\rm i}(k)/L$ with the single-particle dispersion $\varepsilon_k$ of the 
given model, the momentum distribution function in the initial state $n_{\rm i}(k)$, 
and $x$ being the position on the line or a discrete lattice site. 
The $t$ independence is particular to the one-particle Green function; other 
observables and correlation functions depend on $t$. 
Fourier transformation leads to $\rho_{\rm ss}^<(\omega) = \sum_{k} n_{\rm i}(k) 
\delta(\omega-\varepsilon_k)/L$. The complexity of computing
$\rho_{\rm ss}^<(\omega)$ for the inverse quench is thus comparable to the one of computing 
$n_{\rm i}(k)$ in the interacting ground state of microscopic models from the LL class. 
The latter was recently achieved with high precision  
using DMRG \cite{Karrasch12a}. As $\rho_{\rm ss}^<(\omega)$ `samples' the equilibrium 
$n_{\rm i}(k)$ and for {\em any model} from the LL class $|1/2-n_{\rm i}(k)| 
\sim |k-k_F|^{\gamma_{\rm gs}}$ \cite{Giamarchi03,Schoenhammer05} 
it is evident that the main characteristics of $\rho_{\rm ss}^<(\omega)$ are the same as 
the ones discussed above: a finite weight at $\omega=0$ which is approached for 
$\omega \to 0^\pm$ following a power law with interaction dependent 
exponent (for $\gamma_{\rm gs} <1$; 
else $\rho_{\rm ss}^<(\omega) - \rho_{\rm ss}^<(0) \sim -\omega$). This similarly holds for 
$\rho_{\rm ss}^<(k,\omega) \sim  n_{\rm i}(k) \delta(\omega-\varepsilon_k)$ in which 
a positive frequency peak shows up for $k-k_{\rm F}>0$ as $n_{\rm i}(k > k_{\rm F})$ 
takes a nonvanishing value (LL property).     

In Fig.~\ref{fig3}(a) we show $\rho^<_{\rm ss}(\omega)$ after the inverse 
interaction quench for the lattice model of spinless fermions with nearest 
neighbor interaction $H=-J \sum_j c_j^\dag c_{j+1}^{} + {\rm H.c.} + U n_j n_{j+1}$, 
with $n_j = c_j^\dag c_{j}^{}-1/2$, at half filling. 
We did not use $n_{\rm i}(k)$ but equally efficiently computed 
$\rho_{\rm ss}^<(\omega)$ by time-dependent (T)DMRG \cite{Schollwoeck11} (in $t'$ at $t=0$ relying on 
the $t$ independence) and numerical Fourier transformation exploiting that the time evolution 
is a {\em free} one \cite{SM}. This allows us to reach times $t'J=\mathcal{O}(10^3)$ 
much larger than usual in TDMRG. 

We finally study the inverse quench for our lattice model assuming {\em open
boundaries.} Then $G_t^<(j|j',t')$ {\em depends on} $t$ (as well as on $j$ and $j'$ 
and not only $j-j'$). Exploiting again that the time evolution is a free one 
using TDMRG we can compute $G_t^<(j|j,t')$ at finite $t$ for $t'$ so large that
Fourier transformation is meaningful. For $t \to \infty$ the steady state 
local spectral function $\rho_{\rm ss}^<(j,\omega)$ can be accessed computing the 
average of $c_j^\dag(0) c_j^{}(t')$ with the GGE density matrix  
$\tilde \rho_{\rm GGE} = \exp{(-\sum_{k} \tilde \lambda_k c_k^\dag c_k^{})}
/\tilde Z_{\rm GGE}$ and the Lagrange multipliers $\tilde \lambda_k$ fixed 
such that $\left< {\rm gs} \right| n_k \left| {\rm gs}\right> = 
\left< n_k\right>_{\tilde \rho_{\rm GGE}}$ with $n_k=c_k^\dag c_k^{}$ and 
$c_k = \sqrt{2/(L+1)} \sum_j \sin(kj) c_j$. 
For the time evolution with an interaction free Hamiltonian this was shown 
explicitly in Ref.~\cite{Sotiriadis14}. 
This leads to $\rho_{\rm ss}^<(j,\omega)= 
2 \sum_{k} \sin^2(kj) n_{\rm i}(k) \delta(\omega - \varepsilon_k)/(L+1)$, with the 
initial `pseudo-momentum' distribution function $ n_{\rm i}(k) = \left< {\rm gs} \right| 
n_k \left| {\rm gs} \right>$. The latter can be computed using DMRG \cite{SM}. 
From `open boundary' bosonization of the TLM \cite{Giamarchi03,Schoenhammer05} we
expect $|1/2-n_{\rm i}(k)| \sim |k-k_F|^{\gamma_{\rm gs}}$ 
(for $\gamma_{\rm gs}<1$;  else $1/2-n_{\rm i}(k) \sim k-k_F$) which implies 
that $\rho_{\rm ss}^<(j,\omega)$ shows the same spectral features as $\rho_{\rm ss}^<(\omega)$ 
of the translational invariant case. This is confirmed in Fig.~\ref{fig3}(b) in which the 
evolution of $\rho_{t}^<(j=1,\omega)$ towards $\rho_{\rm ss}^<(j=1,\omega)$ is shown. We note 
that only $\rho_{\rm ss}^<$, not $\rho_{t}^<$ at finite $t$, can be 
measured in `continuous beam' photoemission spectroscopy \cite{Freericks09}; $\rho_t^<$
is merely an auxiliary quantity. 

In summary, we revealed spectral features which are unique to the intermediate time 
nonequilibrium LL steady state reached after a global interaction quench. They can be used 
to identify this state in quenched cold Fermi gases in future radio-frequency spectroscopy.

\textit{Acknowledgments}---We thank T.~Enss, F.~Heidrich-Meisner, S.~Jakobs, and C. Karrasch 
for discussions.

{}


\begin{thebibliography}{}

\bibitem{Cazalilla06} 
  M.A.~Cazalilla, Phys.~Rev.~Lett. {\bf 97},  156403 (2006).

\bibitem{Rigol07}
  M.~Rigol, V.~Dunjko, V.~Yurovsky, and M.~Olshanii, Phys.~Rev.~Lett. {\bf 98}, 050405 (2007).

\bibitem{Kollath07} C. Kollath, A. M. L\"auchli, and E. Altman, Phys. Rev. Lett. {\bf 98}, 180601 (2007).

\bibitem{Manmana07}
  S.R.~Manmana, S.~Wessel, R.M.~Noack, and A.~Muramatsu, Phys.~Rev.~Lett. {\bf 98}, 
  210405 (2007). 

\bibitem{Kollar08}
  M.~Kollar and M.~Eckstein, Phys.~Rev.~A {\bf 78} 013626 (2008).

\bibitem{Barthel08} 
  T.~Barthel and U.~Schollw\"ock, Phys.~Rev.~Lett. {\bf 100},  100601 (2008).

\bibitem{Mitra11}
  A.~Mitra and T.~Giamarchi, Phys.~Rev.~Lett. {\bf 107}, 150602 (2011). 

\bibitem{Karrasch12} C. Karrasch, J. Rentrop, D. Schuricht, and V. Meden, Phys. Rev. Lett. 
{\bf 109}, 126406 (2012). 

\bibitem{Mitra12} 
  A.~Mitra and T.~Giamarchi, Phys.~Rev.~B {\bf 85}, 075117 (2012). 

\bibitem{Mitra13} 
  A.~Mitra, Phys.~Rev.~B {\bf 87}, 205109 (2013).

\bibitem{Sirker14}
  J. Sirker, N.P. Konstantinidis, F. Andraschko, N. Sedlmayr, 
  Phys. Rev. A {\bf 89}, 042104 (2014). 

\bibitem{Essler14} 
  F.H.L. Essler, S. Kehrein, S.R. Manmana, and N.J. Robinson, 
  Phys. Rev. B {\bf 89}, 165104 (2014). 

\bibitem{Bloch08}
  I.~Bloch, J.~Dalibard, and W.~Zwerger, Rev.~Mod.~Phys. {\bf 80}, 885 (2008).

\bibitem{Polkovnikov11}
  A.~Polkovnikov, K.~Sengupta, A.~Silva, and M.~Vengalattore, Rev.~Mod.~Phys. 
  {\bf 83}, 863 (2011).

\bibitem{Kennes13}
  D. M. Kennes and V. Meden,
  Phys. Rev. B {\bf 88}, 165131 (2013).

\bibitem{Iucci09}
  A.~Iucci and M.A.~Cazalilla, Phys.~Rev.~A {\bf 80}, 063619 (2009).

\bibitem{Kennes10}
  D.M. Kennes and V. Meden, Phys.~Rev. B {\bf 82}, 085109 (2010). 

\bibitem{Calabrese11}
  P. Calabrese, F.H.L. Essler, and M. Fagotti, Phys.~Rev.~Lett. {\bf 106}, 227203 (2011).  

\bibitem{Essler12}
  F.H.L. Essler, S. Evangelisti, and M. Fagotti, Phys. Rev. Lett. {\bf 109}, 247206 (2012).

\bibitem{Fagotti13} 
  M. Fagotti and F.H.L. Essler, Phys. Rev. B {\bf 87}, 245107 (2013). 

\bibitem{Sotiriadis14} 
  S. Sotiriadis and P. Calabrese, , J. Stat. Mech. P07024 (2014).

\bibitem{Fagotti14}
  M. Fagotti, M. Collura, F.H.L. Essler, and P. Calabrese,
  Phys. Rev. B {\bf 89}, 125101 (2014). 
  
\bibitem{Wouters14}
B. Wouters, M. Brockmann, J. De Nardis, D. Fioretto, and J.-S. Caux, Phys. Rev. Lett. {\bf 113}, 117202 (2014).



\bibitem{Mierzejewski14}
M. Mierzejewski, P. Prelov\u{s}ek, T. Prosen, Phys. Rev. Lett. {\bf 113}, 020602 (2014).

\bibitem{Pozsgay14}
B. Pozsgay, M. Mesty\'an, M.A. Werner, M. Kormos, G. Zar\'and, and G. Tak\'acs, Phys. Rev. Lett. {\bf 113}, 117203 (2014).
.

\bibitem{Giamarchi03}
  T. Giamarchi, {\em Quantum Physics in One Dimension} (New York: Oxford 
  University Press, 2003).

\bibitem{Schoenhammer05} 
  K. Sch\"onhammer in
  {\em Interacting Electrons in Low Dimensions} ed. by D. Baeriswyl (Dordrecht: 
  Kluwer Academic Publishers, 2005). 
    
\bibitem{Blumenstein11}
  C. Blumenstein, J. Sch\"afer, S. Mietke, S. Meyer, A. Dollinger, M. Lochner, 
  X.Y. Cui, L. Patthey, R. Matzdorf, and R. Claessen,
  Nature Physics {\bf 7}, 776 (2011).

\bibitem{Uhrig09} G.S. Uhrig, Phys. Rev. A {\bf 80}, 061602 (2009).

\bibitem{Rentrop12}
  J. Rentrop, D. Schuricht, and V. Meden,
  New J. Phys. {\bf 14}, 075001 (2012).

 \bibitem{Schollwoeck11} U.~Schollw\"ock, Ann. Phys. {\bf 326}, 96 (2011).  
  
\bibitem{Stewart08} 
  J.T. Stewart, J.P. Gaebler, and D.S. Jin,
  Nature {\bf 454}, 744 (2008).

\bibitem{Gaebler10}
  J.P. Gaebler, J.T. Stewart, T.E. Drake, D.S. Jin, A. Perali, P. Pieri, and G.C. Strinati,
  Nature Physics {\bf 6}, 569 (2010).

\bibitem{Feld11}
  M. Feld, B. Fr\"ohlich, E. Vogt, M. Koschorreck, and M. K\"ohl, Nature {\bf 480}, 75, (2011). 

\bibitem{SM} See Supplemental Material [url], which includes Refs. \onlinecite{Meden92,Barthel09,Hartmann09,Muth11,Vidal,Kennes14,Jeckelmann08}. 

\bibitem{Meden92}
  V. Meden and K. Sch\"onhammer, 
  Phys. Rev. B {\bf 46}, 15753 (1992). 

 \bibitem{Barthel09}
 T.~Barthel, U.~Schollw\"ock and S.R.~White,  Phys. Rev. B {\bf 79}, 245101 (2009). 

 \bibitem{Hartmann09} 
 M.J. Hartmann, J. Prior, S.R. Clark and M.B. Plenio, Phys. Rev.Lett. {\bf 102}, 057202 (2009).  

  \bibitem{Muth11}
D. Muth and  R.G. Unanyan, and M. Fleischhauer, Phys. Rev. Lett. {\bf 106}, 077202 (2011).
  
  \bibitem{Vidal}  G. Vidal, Phys. Rev. Lett. {\bf 98}, 070201 (2007). 
  
 \bibitem{Kennes14}  D. M. Kennes and C. Karrasch, arXiv:1404.3704 (2014). 
 
  
  \bibitem{Jeckelmann08}
  E. Jeckelmann and H. Benthien in {\em Computational Many Particle Physics,} H. Fehske, 
  R. Schneider, and A. Weiße (Eds.), Lecture Notes in Physics 739, pp. 621-635, 
  Springer-Verlag, Berlin, Heidelberg, (2008).
  

\bibitem{Meden99}
  V.~Meden, Phys.~Rev.~B {\bf 60}, 4571 (1999).  

\bibitem{Schoenhammer93} 
  K. Sch\"onhammer and V. Meden, 
  J. Electron. Spectrosc. {\bf 62}, 225 (1993). 

\bibitem{Karrasch12a} 
  C. Karrasch and J. E. Moore, 
  Phys. Rev. B {\bf 86}, 155156 (2012).
  
\bibitem{Freericks09}  
  J.K. Freericks, H.R. Krishnamurthy, and Th. Pruschke,
  Phys. Rev. Lett. {\bf 102}, 136401 (2009). 
  



  
\end{thebibliography}
\end{document}


\title{Supplementary material for \\ `Spectral properties of one-dimensional Fermi 
systems after an interaction quench'}

\author{D.M.\ Kennes}  
\affiliation{Institut f{\"u}r Theorie der Statistischen Physik, RWTH Aachen University 
and JARA---Fundamentals of Future Information
Technology, 52056 Aachen, Germany}

\author{C.\ Kl\"ockner}  
\affiliation{Institut f{\"u}r Theorie der Statistischen Physik, RWTH Aachen University 
and JARA---Fundamentals of Future Information
Technology, 52056 Aachen, Germany}

\author{V.\ Meden} 
\affiliation{Institut f{\"u}r Theorie der Statistischen Physik, RWTH Aachen University and 
JARA---Fundamentals of Future Information
Technology, 52056 Aachen, Germany}
 
 


\maketitle

\begin{widetext}

\vspace{-1.cm}

\section{Exact analytical expressions for the lesser Green function}

We here present the analytical expressions for the lesser Green function $G^<_t(x,t')$ 
of the TLM A. in the ground state, B. after the global quench of the two-particle interaction from
zero to $g(q)$, and C. in the equilibrium canonical ensemble at temperature 
$T$. They are obtained by standard bosonization of the 
field operator [\onlinecite{Giamarchi03},\onlinecite{Schoenhammer05}]. 
As outlined in the main text when computing $n(k)$ and $\rho^<(\omega)$ 
we do not rely on the commonly used approximate ad hoc procedure which allows to 
analytically evaluate the 
remaining momentum integral in $G^<_t(x,t')$ [\onlinecite{Meden99}]. 

\subsection{Ground state}

The $t$ independent lesser Green function can be expressed in terms of the noninteracting one
$G^<_0(x,t')$ and an interaction dependent function $F(x,t')$ as
\begin{eqnarray}
i G^<_t(x,t') = i G^<_0(x,t') e^{F(x,t')} .
\label{form}
\end{eqnarray}
The former is given by
\begin{eqnarray}
i G^<_0(x,t') = - \frac{i}{2 \pi} \, \frac{e^{i k_{\rm F} x}}{x- v_{\rm F} t'- i \eta} ,
\label{G0}
\end{eqnarray}
where we have already taken the thermodynamic limit and $\eta$ must be sent to $0^+$. 
The interaction enters via
\begin{eqnarray}
F_{\rm gs} (x,t') = \int_0^\infty \frac{dq}{q} \, \left\{  e^{-i[qx-\omega(q) t']} -  
e^{-i[qx-v_{\rm F} q t']} + 2 s^2(q) \left[  \cos(qx) e^{i \omega(q) t'} -1 
\right] \right\} ,
\end{eqnarray}
with $\omega(q)$ and $s^2(q)$ as given in the main text. We note that the $q$ integral 
is cut off in the ultraviolet by the range of the two-particle potential in momentum space 
$q_{\rm c}$. This can explicitly be seen in the third term proportional to $s^2(q)$ 
as the latter approaches zero on the scale $q_{\rm c}$.  It is less obvious 
for the first two terms, which, however, cancel each other for $q/q_{\rm c} \gg 1$ as 
$\omega(q) \to v_{\rm F} q$ in this limit. Thus no further regularization is required. 
The same holds for the cases B. and C. discussed below.  

Setting $t'=0$ and Fourier transforming leads to
\begin{eqnarray}
\left[n_{\rm gs}(k) -\frac{1}{2} \right] \, {\rm sgn}(k_{\rm F}-k) = \int_{-\infty}^{\infty} \frac{dx}{2 \pi} 
\, \frac{\sin(|k-k_{\rm F}|x)}{x} e^{F_{\rm gs}(x,0)} .
\label{nk}
\end{eqnarray}
The remaining two nested integrals can easily be performed numerically. Furthermore,
this expression forms the basis to prove $[n_{\rm gs}(k) -\frac{1}{2}] \, {\rm sgn}(k_{\rm F}-k) 
\sim |k-k_{\rm F}|^{\gamma_{\rm gs}}$ for $\gamma_{\rm gs} <1$ and  $n_{\rm gs}(k) -\frac{1}{2} 
\sim k_{\rm F} - k$ for $ \gamma_{\rm gs} > 1$  using asymptotic analysis. 
For $\gamma_{\rm gs}=1$ one finds a logarithmic corrections to the linear term [\onlinecite{Meden99}].

As mentioned in the main text $F_{\rm gs}(0,t')$ and thus $G^<_t(0,t')$ is analytical in the upper 
half of the complex $t'$ plane. This implies $\rho^<_{\rm gs} \sim \Theta(-\omega)$. More specifically
we find
\begin{eqnarray}
2 \pi v_{\rm F} \rho^<_{\rm gs}(\omega) = \Theta(-\omega) \left\{ \frac{1}{2 } - \frac{1}{2\pi} \, \int_{-\infty}^{\infty} 
\!\!\!\!\!\! \!\!\!\!\!\!\! \mathcal{P} \;\;\;\;\; \frac{dt'}{t'} \, {\rm Im} \left[ e^{i \omega t'} 
e^{F_{\rm gs}(0.t')}  \right]      \right\}
\label{rhoomega}
\end{eqnarray}
Again the two remaining nested integrals can be performed numerically and Eq.~(\ref{rhoomega}) is the basis 
to prove  $\rho^<_{\rm gs}(\omega) \sim (-\omega)^{\gamma_{\rm gs}}$  (for all $\gamma_{\rm gs}$) 
using asymptotic analysis.   

\subsection{Quench dynamics and steady state}

We next give the $t$ dependent lesser Green after the interaction quench. It can also be written as 
the product Eq.~(\ref{form}) with the noninteracting Green function Eq.~(\ref{G0}) and
\begin{eqnarray}
F_{t} (x,t')  & =  & \int_0^\infty  \frac{dq}{q} \, \left\{ 
-4 s^2(q) c^2(q) + 2 c^2(q) s^2(q) \left[ \cos(2 \omega(q) t)  +  
\cos(2 \omega(q) [t+t'])  \right] \right. \nonumber \\
&&  + e^{-iqx} \left[ c^4(q) e^{i \omega(q) t'} - e^{i v_{\rm F} q t'} 
+s^4(q) e^{-i \omega(q) t'} 
-2 c^2(q) s^2(q) \cos(\omega(q) [2 t+t'])
\right] \nonumber \\
&& \left. + 2 e^{iqx} c^2(q) s^2(q)  \left[ \cos( \omega(q) t') - \cos(\omega(q) [2 t+t'])
\right]
\right\} ,
\end{eqnarray}
with $c^2(q) = 1+s^2(q)$.
For large times $t$ after the quench all oscillatory terms containing $t$ drop out when 
performing the $q$ integral 
and we obtain  
\begin{eqnarray}
F_{\rm ss} (x,t')  =   \int_0^\infty  \!\! \frac{dq}{q} \left\{ 
-4 s^2(q) c^2(q) 
+ e^{-iqx} \left[ c^4(q) e^{i \omega(q) t'} - e^{i v_{\rm F} q t'} 
+s^4(q) e^{-i \omega(q) t'} \right] + 2 e^{iqx} c^2(q) s^2(q) \cos[ \omega(q) t'] 
\right\} .
\label{Fss}
\end{eqnarray}
As mentioned in the main text this expression can directly be derived 
using a GGE [\onlinecite{Essler12}]. To achieve this one takes the expectation value 
$\left< \psi^{\dag}(0,0) \psi^{}(x,t')\right>_{\rho_{\rm GGE}}$ with the GGE 
density matrix $\rho_{\rm GGE} = \exp{(-\sum_{n \neq 0} \lambda_n \alpha_n^\dag \alpha_n^{})}/Z_{\rm GGE}$. 
The Lagrange multipliers $\lambda_n$ are fixed such that $\left< {\rm gs}_0 \right| N_n 
\left| {\rm gs}_0 \right> = \left< N_n\right>_{\rho_{\rm GGE}}$ 
for the conserved eigenmode occupancies $N_n = \alpha_n^{\dag} \alpha_n^{}$ 
which leads to $s^2(q_n) = \left[ e^{\lambda_n} -1 \right]^{-1}$.  
Finally, one has to take  $L \to \infty$. In contrast to $F_{\rm gs}(0,t')$, $F_{\rm ss}(0,t')$
is not analytical in the upper half of the complex $t'$ plane and $\rho^<_{\rm ss}(\omega)$ has finite 
weight for all $\omega$. 

Based on Eqs.~(\ref{form}) and (\ref{Fss}) $n_{\rm ss}(k)$ and $\rho_{\rm ss}(\omega)$ can be brought 
in forms similar to Eqs.~(\ref{nk}) and (\ref{rhoomega}), respectively, with $F_{\rm gs}$ replaced by
$F_{\rm ss}$. In each case the remaining two nested integrals can be performed numerically. 
Furthermore, the respective expressions are the basis to prove $[n_{\rm ss}(k) -\frac{1}{2}] \, 
{\rm sgn}(k_{\rm F}-k) \sim |k-k_{\rm F}|^{\gamma_{\rm ss}}$ ($\gamma_{\rm ss} < 1$; for the 
case $\gamma_{\rm ss} \geq 1$ the same as for $n_{\rm gs}(k)$ holds) as well as 
$|\rho^<_{\rm ss}(\omega) - \rho^<_{\rm ss}(0)| \sim |\omega|^{\gamma_{\rm ss}}$ for 
$\gamma_{\rm ss}<1$ and 
$\rho^<_{\rm ss}(\omega) - \rho^<_{\rm ss}(0) \sim - \omega$ for 
$\gamma_{\rm ss}>1$.

\subsection{Canonical ensemble}

Also the $t$ independent equilibrium lesser Green function in the thermal canonical ensemble can be 
written in the form Eq.~(\ref{form}). This time the noninteracting Green function is given by
\begin{eqnarray}
i G^<_0(x,t') = - \frac{i}{2 \pi} \, e^{i k_{\rm F} x}\, \frac{\pi T/v_{\rm F}}{
\sinh\left[ (x- v_{\rm F} t'- i \eta) \pi T/v_{\rm F}\right]} 
\label{G0th}
\end{eqnarray}
and the interaction dependent part by
\begin{eqnarray}
F_{\rm th} (x,t') \!\! & = & \!\! \int_0^\infty \!\! \frac{dq}{q} \!\! \left\{  e^{-i[qx-\omega(q) t']} -  
e^{-i[qx-v_{\rm F} q t']} 
- 2 b(\omega(q)/T) \left( 1- \cos[qx-\omega(q) t'] \right)
+ 2 b(v_{\rm F} q/T) \left( 1- \cos[qx-v_{\rm F} q t'] \right) \right. \nonumber  \\
&& \left. + 2 s^2(q) \left[  \cos(qx) \left(  e^{i \omega(q) t'} [1+b(\omega(q)/T)] 
+ e^{-i \omega(q) t'} b(\omega(q)/T) \right) - \left( 1 + 2 b(\omega(q)/T)  \right)  \right] 
\right\} 
\end{eqnarray}
with $b(x)=\left( e^{x}-1 \right)^{-1}$.
For $T>0$, $F_{\rm th}(0,t')$ is no longer analytical in the upper half of the complex $t'$ 
plane. Thus $\rho^<_{\rm th}(\omega)$ has nonvanishing weight also for $\omega>0$. 
The nested Fourier and momentum integrals leading to $n_{\rm th}(k)$ and $\rho^<_{\rm th}(\omega)$ 
can be performed numerically. In the respective Fourier integrals we take 
$\eta q_{\rm c}=10^{-3}$. This weakly affects the momentum distribution function and momentum 
integrated spectral function at large $|k-k_{\rm F}|$ and $|\omega|$, respectively. 
However, for the arguments shown in the figures of the main text further reducing 
$\eta$ does not change the graphs on the respective scales. In [\onlinecite{Schoenhammer93}] it 
is discussed how the limiting 
behavior of $\rho^<_{\rm th}(\omega)$ at $|\omega|/(v_{\rm F} q_{\rm c}), T/(v_{\rm F} q_{\rm c})  
\ll 1$ mentioned in the main text can be obtained.

\section{The ad hoc procedure}

To avoid the numerical evaluation of three nested integrals when computing 
$\rho^<(k,\omega)$ we resort to an often used ad hoc procedure which results 
in {\em approximate} but integral-free expressions of the three 
$G_t^<(x,t')$ of interest (see [\onlinecite{Meden99}] for the ground state). 
The consequences of the underlying approximations for the momentum resolved spectral 
function are mentioned in the main text. The procedure consists of several steps. 
In a first one starting from Eq.~(\ref{form}) the noninteracting Green function is 
rewritten as a momentum integral or, at finite length $L$, as a momentum sum. We here 
consider the general finite $T$ case, from which the ground state one relevant for 
the ground state Green function as well as the one after the quench, can easily be 
obtained by taking $T \to 0$. Written as a sum $G_{0}^<(x,t')$ reads
\begin{eqnarray}
i G_{0}^<(x,t') = \frac{e^{i k_{\rm F} x}}{L} \, \exp \left\{ \sum_{n=1}^\infty \frac{1}{n}
\left[  e^{i v_{\rm F} q_n t'} - 2 b(v_{\rm F}q_n/T) \left( 1- \cos\left[ v_{\rm F} q_n t'\right] \right)  
\right]\right\}. 
\label{G0adhoc}
\end{eqnarray}
The terms appearing in the argument of the exponential function cancel corresponding ones in 
the expressions for the three $F$'s if those are written for finite $L$, that is as sums. Next 
in the respective $F$ one 
linearizes $\omega(q) = v q$, with $v=v_{\rm F} \sqrt{1+\hat g(0)}$, for all $q$ 
which constitutes the main approximation inherent to the ad hoc procedure. Now terms similar 
to the ones on the right-hand-side of Eq.~(\ref{G0adhoc}) appear, but with $v_{\rm F}$ 
replaced by $v$. For those the momentum sums are performed in the same way which 
leads from Eq.~(\ref{G0adhoc}) to Eq.~(\ref{G0th}). Finally, in the remaining 
$q$ sum appearing in the different $F$'s one replaces $s^2(q)$ by $s^2(0) \exp(- |q/q_{\rm c}|)$---as 
well as $c^2(q)$ by $1+s^2(0) \exp(- |q/q_{\rm c}|)$ in the steady state---and this way 
regularizes the sum in the ultraviolet `by hand'. All momentum sums can then be performed 
analytically. After this procedure $v_{\rm F}$ completely dropped out 
which explains why in Fig. 2 of the main text $v$ instead of $v_{\rm F}$ appear in the scales 
of both the $x$ as well as $y$ axis. 

In the thermodynamic limit the closed approximate expressions for the Green functions obtained
along the given line read 
\begin{eqnarray}
i G_{\rm gs}(x,t') \approx - \frac{i}{2 \pi} \, \frac{e^{i k_{\rm F} x}}{x- v t'- i \eta} 
\left[ \frac{r^2}{(x - v t' -i r) (x+vt'+ir)}   \right]^{s^2(0)},
\end{eqnarray}
with $r=q_{\rm c}^{-1}$, for the ground state [\onlinecite{Meden99}],
\begin{eqnarray}
i G_{\rm ss}(x,t') & \approx & - \frac{i}{2 \pi} \, \frac{e^{i k_{\rm F} x}}{x- v t'- i \eta} 
\left[ \frac{r}{x - v t' - i r }   \right]^{2 s^2(0)}
\left[ \frac{r^2}{(x - v t' +i r) (x+vt'+ir)}   \right]^{s^2(0)} \nonumber \\
&& \times \left[ \frac{(2r)^4}{(x + v t' -i 2 r) (x-vt'+i2r)  (x - v t' -i 2 r) (x+vt'+i2r) }   \right]^{s^4(0)} . 
\end{eqnarray}
for the steady state after the interaction quench, and
\begin{eqnarray}
i G^<_{\rm th}(x,t') & \approx & - \frac{i}{2 \pi} \, e^{i k_{\rm F} x}\, \frac{\pi T/v}{
\sinh\left[ (x- v t'- i \eta) \pi T/v\right]} 
\left[ \frac{r^2}{(x - v t' -i r) (x+vt'+ir)}   \right]^{s^2(0)} \nonumber \\
&& \times \left[ \frac{\Gamma(a+i u_+) \Gamma(a-i u_+) \Gamma(a+i u_-) \Gamma(a-i u_-)}{\Gamma^4(a)} \right]^{s^2(0)} ,
\end{eqnarray}
with $a=1+T/(v q_{\rm c})$, $u_\pm =T (x \pm vt')/v$, and the Gamma function $\Gamma$, for 
the finite $T$ equilibrium Green function [\onlinecite{Schoenhammer93}]. 

In the computation of 
\begin{eqnarray}
\rho^<(k,\omega) = \frac{1}{2 \pi} \int_{-\infty}^\infty dx \int_{-\infty}^\infty
dt'  e^{-i(kx-\omega t')}  i G^<(x,t')
\end{eqnarray}
based on the approximate $G^<(x,t')$ we substitute $z_\pm=x \pm v t$. 
This way the two nested Fourier integrals factorize in all three cases. For $G^<_{\rm gs}(x,t')$ this 
results in a closed analytical expression for the momentum resolved spectral function which 
involves the confluent hypergeometric function; for details see [\onlinecite{Meden92}]. In the other 
cases the integrals can be performed numerically again setting $\eta q_{\rm c}=10^{-3}$ (with minor 
effect on the spectra shown in the figures of the main text). The factorized form also constitutes 
the basis to show that in the ground state for $k<k_{\rm F}$,
$\rho^<_{\rm gs}(k, \omega \approx  v [k-k_{\rm F}]) \sim \Theta(-\omega + v [k-k_{\rm F}]) 
(-\omega + v [k-k_{\rm F}])^{\gamma_{\rm gs}/2-1}$
and for $k>k_{\rm F}$, $\rho^<_{\rm gs}(k, \omega \approx - v [k-k_{\rm F}]) \sim 
 \Theta(-\omega - v [k-k_{\rm F}])  (-\omega - v [k-k_{\rm F}])^{\gamma_{\rm gs}/2}$ using asymptotic analysis. 
Here $\gamma_{\rm gs} = 2 s^2(0)=(K+K^{-1}-2)/2$, with the LL parameter $K^{-1}=\sqrt{1+\hat g(0)}$.
Similar power laws, but without the corresponding $\Theta$-functions, are found in the steady state. 
They read $\rho^<_{\rm ss}(k, \omega \approx  v [k-k_{\rm F}]) \sim  |-\omega + v [k-k_{\rm F}]|^{s^2(0)+2s^4(0)-1}$
(for $s^2(0)+2s^4(0) < 1$) and $|\rho^<_{\rm ss}(k, \omega \approx - v [k-k_{\rm F}]) - 
\rho^<_{\rm ss}(k, \omega = - v [k-k_{\rm F}])| \sim |-\omega - v [k-k_{\rm F}]|^{3 s^2(0) + 2 s^4(0)}$
(for $3 s^2(0) + 2 s^4(0)<1$). Using
the above relation between $s^2(0)$ and $K$ these exponents can be written in terms of $\gamma_{\rm ss} = 
(K^2+K^{-2}-2)/4$, which, however, is not very instructive.

\section{DMRG for the lattice model}

We use DMRG [\onlinecite{Schollwoeck11}] to compute $\rho^<_t(j,\omega)$ for the lattice model described by 
\begin{equation}
H=-J\sum\limits_{j}c_j^\dagger c_{j+1}+{\rm H.c.}+Un_jn_{j+1},
\label{eq:microH}
\end{equation}
with $n_j=c_j^\dagger c_j-1/2$. $\rho^<_t(j,\omega)$ is the Fourier transform with respect to $t'$ 
of the correlator 
\begin{equation}
G^<_t(j,t')=\left \langle\psi(t)\right|c^\dagger_j e^{iHt'}c_{j}e^{-iHt'}\left|\psi(t)\right\rangle\;,\;\;\;\;
\;\left|\psi(t)\right\rangle=e^{-iHt}\left|\psi(0)\right\rangle,
\label{eq:Gless_micro}
\end{equation}
which  can be implemented straightforwardly by first propagating 
$\left|\psi(0)\right\rangle$ to $\left|\psi(t)\right\rangle$ and 
then calculating the overlap of $e^{-iHt'}c_{j}^\dagger\left|\psi(t)\right\rangle$ 
and $c_j e^{-iHt'}\left|\psi(t)\right\rangle$. 

However, for a general global quench the bond dimension of the DMRG procedure 
determining the numerical effort grows exponentially in time. This severly restricts 
the times reachable which in turn results in a low energy resolution after Fourier transforming. 
We are partly interested in the power-law behavior found in the spectral function for small 
frequencies. We thus do not want to use tools like linear prediction, which allows to extrapolate 
data to larger times thus implying improved energy resolution which, however, is only partly 
controlled [\onlinecite{Barthel09}]. Therefore, we focus on the quench protocol in which 
the ground state $\left|{\rm gs}\right\rangle$ with respect to the full 
Hamiltonian Eq. \eqref{eq:microH} is prepared and than the interaction  $U$ is 
suddenly switched off. We now change perspective and propagate the operators 
$c^\dagger_j$ and $c_j$ (calculating $e^{iHt}c^\dagger_j e^{iHt'}c_{j}e^{-iH(t'+t)}$) 
instead of the wave function's matrix product state (MPS). It was shown, that the 
time evolution with the free Hamiltonian can be done \emph{exactly} with finite 
bond dimension equal to four [\onlinecite{Hartmann09},\onlinecite{Muth11}]. 
The numerical effort to perform the time evolution reduces from exponential 
to linear growth in time and times $Jt\sim \mathcal{O}(10^3)$ become easily accessible. 

For Fig.~3(a) of the main text showing results for the translational invariant model we 
used this approach to calculate the correlator Eq.~\eqref{eq:Gless_micro} at $t=0$, 
where the MPS was prepared by an infinite DMRG ground state search [\onlinecite{Vidal}]. 
After drawing a random MPS an imaginary time evolution with decreasing trotter step 
size ($\Delta \tau\in\{0.2,0.02,0.002,0.0002,0.00002,0.000002\}$) was performed until 
for each of these step sizes the energy was converged to the first $10$ significant digits. 
For our problem a bond dimension $\chi= 200$ was found to be sufficient to achieve 
converged results. For the matrix product operator (MPO) [\onlinecite{Kennes14}]  of $c_j^{(\dagger)}$, which 
we propagate in time, we set the length of the operator 
$1\otimes 1\otimes \dots c_j^{(\dagger)}\dots \otimes 1$ to $500$ lattice sites and 
abort the propagation in time before the effects of the finite length are visible. 
As we can time evolve the operator with bond dimension $\chi=4$ no truncation is 
needed. With this, our approach is free of any finite size effects. 

Analogously, but with a finite system of $L=2000$ lattice sites and open boundaries we 
computed the curves for finite $t$ shown Fig.~4(b). Here, instead of the infinite 
DMRG procedure, we employed an iterative ground state search [\onlinecite{Schollwoeck11}] with 
$20$ sweeps in each direction. We compared different discarded weights 
$\epsilon\in\{10^{-6},10^{-7},10^{-8}\}$ to monitor convergence. 

In the model with open boundaries the limit $t\to \infty$ can be addressed explicitly 
by calculating $n_{\rm i}(k)$ of the interacting ground state 
[\onlinecite{Karrasch12a},\onlinecite{Jeckelmann08}] and using  
\begin{align}
\rho(j,\omega)&=\frac{2}{L+1}\sum\limits_{n=1}^L\sin^2(k_n j)n_{\rm i}(k_n)\delta(\omega-\epsilon_{k_n})
\overset{L \to \infty}{\longrightarrow}   \frac{2}{\pi} \frac{1}{|2\sin(k^*(\omega))|}W_j(k^*(\omega)), \label{eq:rho}\\
k^*(\omega)&={\rm arcos}(-\omega/2),\\
W_j(k)&=\sin^2(k j)n_{\rm i}(k),\\
n_{\rm i}(k_n)&=\frac{2}{L+1}\sum\limits_{j,j'=1}^L\sin(k_nj)\sin(k_nj')\left\langle {\rm gs}
\right|c^\dagger_jc_{j'}\left|{\rm gs}\right\rangle\\
k_n&= \frac{\pi}{L+1} n ,\;\;\;\;\; n = 1,2,\ldots,L.
\end{align}
More specifically we computed the matrix elements $\left\langle {\rm gs}
\right|c^\dagger_jc_{j'}\left|{\rm gs}\right\rangle$ by DMRG again considering $L=2000$ 
lattice sites and the same numerical parameters as for the finite $t$ data.
The steady state curves of Fig.~4(b) then follow from this and the above equations 
($\omega$ evaluated at $k^*=k_n$).

\end{widetext}

{}